\documentclass{eas}
\usepackage{graphicx}
\def\doo{d \Omega^2_k} 
\def\S{\Sigma}

\def\Br{\tilde{r}}

\def\Blambda{\tilde{\lambda}}
\def\Bt{{\tilde{t}\,}}

\def\ff{N}

\def\Journal#1#2#3#4#5#6{#5  {#1} {\bf #2} #3}
\def\PRL{\em Phys. Rev. Lett.}
\def\PRD{\em Phys. Rev. D}
\def\CQG{\em Class. Quantum Grav.}
\begin{document}

\title{Accelerating expansion and change of signature} 
\author{Marc Mars}\address{Dept. de F\'{\i}sica Fundamental,
Universidad de Salamanca, Plaza de la Merced s/n, 37008 Salamanca, Spain}
\author{Jos\'e M. M. Senovilla}\address{F\'{\i}sica Te\'orica,
Universidad del Pa\'{\i}s Vasco, Apartado 644, 48080 Bilbao, Spain}
\author{Ra\"ul Vera}\sameaddress{2}
\thanks{MM was supported by the projects FIS2006-
05319 (MEC)  and SA010CO5 (JCyL). JMMS thanks support under grants
FIS2004-01626 (MEC) and GIU06/37 of the University of
the Basque Country.
RV is funded by the Basque
Governement Ref. BFI05.335 and thanks support from
project GIU06/37.}
\begin{abstract}
We show that some types of sudden singularities admit
a natural explanation in terms of regular
changes of signature on brane-worlds in AdS$_{5}$. 
The present accelerated expansion of the Universe and its possible
ending at a sudden singularity may therefore simply be
an indication that our braneworld is about to change its Lorentzian
signature to an Euclidean one, while remaining fully regular. 
An explicit example of this behaviour
satisfying the weak and strong energy conditions is presented.
%
\end{abstract}
\maketitle
\section{Introduction}
As shown in (Mars \etal \/ \cite{sign}) and (Mars \etal \/ \cite{signlong}),
brane-world models are a natural scenario for the \emph{regular}
description of classical signature change.
In brane --and thin shell-- cosmology settings the spacetime
corresponds to an embedded submanifold (the brane or shell) in a
higher dimensional Lorentzian space (bulk).
Smooth submanifolds sitting in a bulk of fixed Lorentzian
signature can have varying causal character, which implies that
its induced metric can undergo a change of signature while remaining
perfectly regular. Even though the
change of signature may appear as a dramatical event from \emph{within
the brane}, both the bulk and the brane remain smooth everywhere. 
In particular, observers living in the brane \emph{but} assuming that
their Universe is Lorentzian everywhere may  
be misled to \emph{interpret}
that a curvature singularity arises precisely at the signature change.

In this short contribution we show that a signature change on the brane
might explain an accelerated expansion of the Universe ending in a
sudden singularity of big-freeze type (Catto\"en \& Visser \cite{CV}),
(Bouhmadi-L\'opez \etal \/ \cite{bigfreeze}), in which the Hubble
function diverges at a finite value of the scale factor,
while keeping the energy density and the rest of physical
variables regular and non-negative everywhere; in particular
without violating the weak or strong energy conditions.

\section{Generic branes in AdS$_5$ bulks}
For concreteness we restrict ourselves to anti-de
Sitter (AdS${}_{5}$) bulks. Choosing suitable coordinates, the metric
is
\[
ds^2=-(k+\lambda^2 r^2)dt^2+(k+\lambda^2 r^2)^{-1}d r^2+ r^2\doo,
\]
where 
$\lambda>0$ is a constant and
$\doo$  is the $3$-dimensional
metric of constant sectional curvature $k=1,0,-1$. 
When $k=0,1$ the ranges of the
non-angular coordinates are $-\infty <t<\infty$ and $r> 0$,
while for $k=-1$ we have 
$r > 1/\lambda$. The 
cosmological constant of AdS$_5$ is  $\Lambda_5  =-6 \lambda^2$.

As proven in  Corollary 2 of (Mars \etal \/ \cite{signlong}),  
branes with a change of signature
require an asymmetric set-up, i.e. a bulk with no $Z_2$ symmetry.
We therefore consider the gluing of a region of
AdS${}_{5}$ with a region of another anti-de Sitter space
$\widetilde{\mbox{AdS}}_5$. All quantities in
$\widetilde{\mbox{AdS}}_5$ will carry an
overtilde, and we assume $\tilde\Lambda_{5} \neq
\Lambda_5$.  For simplicity we consider
branes $\S$ with spherical, plane or hyperboloidal symmetry
so that their parametric form read
(we ignore the angular variables which are merely identified)
$\S : \{t=t(\xi),\,  r=r(\xi)\}$ and
$\tilde{\S} : \{ \Bt=\Bt(\xi),\,  \Br=\Br(\xi) \}$.
The matching requires $r(\xi) = \Br(\xi) \equiv a(\xi) $ so that
the first fundamental form on the brane reads
\begin{equation}
\left.{d s^2}\right|_{\S}= \ff(\xi) d\xi^2 + a^2(\xi) 
d \Omega_{k}^2,
\label{eq:ds2FLRW_N}
\end{equation}
where $N(\xi)$ 
controls the embedding functions $t(\xi)$, $\Bt(\xi)$ via
\begin{equation}
    \dot{t} = \frac{\sigma a}{k + \lambda^2 a^2} \sqrt{ \frac{\dot{a}^2}{a^2}
      - \ff \left ( \frac{k}{a^2} + \lambda^2 \right) } \label{eq:ts}
\end{equation}
and a similar equation for $\tilde t$ in terms of $\Blambda$. 
Here dot stands for $d/d\xi$ and $\sigma$,  $\tilde\sigma$
are two free signs.
$N(\xi)$ and $a(\xi)$ are arbitrary functions 
only restricted to satisfy that
both square roots, in (\ref{eq:ts}) and its tilded version, are real.

The brane metric (\ref{eq:ds2FLRW_N}) changes signature
whenever $N$ changes sign. The 
{\em signature-changing set} of $\S$, denoted by $S$,
is defined to be the collection of all ``instants'' 
$\xi= \xi_s$ 
where $\ff$ becomes, or stops being, zero.
We also define the Lorentzian phase $\S_L$ (where $\ff<0$),
the Euclidean phase $\S_E$ (where $\ff>0$) and the null phase $\S_0$ 
(where $\ff=0$). 
In general, all phases are non-empty for a typical signature-changing brane.

The inherited total
energy-momentum tensor on the brane, $\tau_{\mu\nu}$, 
can be shown to have a vanishing, simple 
eigenvalue, another simple and generically non-zero
eigenvalue 
$\hat \varrho$ and a triple eigenvalue $\hat p$
associated with $\doo$ 
(Mars \etal \/ \cite{signlong}). The energy-momentum
$\tau_{\mu\nu}$, and therefore also $\hat \varrho$ and
$\hat{p}$, are affected by a normalisation freedom depending
on the choice of volume element on the brane and are
\emph{smooth everywhere} on $\S$ (for regular $a(\xi)$ and $\ff(\xi)$).


\subsection{FLRW cosmology in the Lorentzian phase}

On the Lorentzian phase $\S_L$ we can change to the usual cosmic time $T$
with the change of variables
$\dot{T} = \sqrt{-\ff}$ on $\S_L$,  so that the metric becomes 
\begin{equation}
  \left.{d s^2}\right|_{\S_L}= - d T^2 +
  a^2 d \Omega_{k}^2 \, .
  \label{oldRW}
\end{equation}
At the signature changing subset $S\cap \overline{\S_{L}}$, this
Lorentzian metric must show some pathology related to the
fact that the signature in the brane is about to change.
Indeed, since we are assuming a smooth brane, we must have
$a |_S \neq 0$. Moreover, 
$\dot a|_S\neq 0$ because otherwise, from
(\ref{eq:ts}) it would follow $\dot t|_S=\dot{\Bt}|_S=0$ which 
is impossible.
Consequently
$a' =\dot{a}/\sqrt{-N}$ {\em diverges
necessarily}
when approaching the signature-changing set
$S\cap\overline{\S_L}$. Equivalently,
the Hubble function
$H\equiv a'/a$ diverges necessarily at $S\cap\overline{\S_L}$,
where $a$ is finite.
This behaviour cannot be found in pure Lorentzian brane cosmologies,
and is a kinematic characteristic of the so-called big-freeze singularities.

This `singularity' concerns
{\em exclusively} the ``Lorentzianity'' of the brane and
{\it only} appears when the geometry is analyzed from
the {\em inner point
of view} of the Lorentzian part of the brane. The 
``singularity'' arises because the cosmic time is pathological
near the signature change, when it stops existing.
No true singularity is there in the brane. 
As already stressed 
{\em both the bulk and the
brane $\S$ are totally regular everywhere} for regular functions
$N(\xi)$ and $a(\xi)$.

Let us describe the type of singularity that any observers
living on $\S_L$ would {\em believe to see} there.
If the scientists on $\S_L$ know the spacetime is  a 4-d brane
immersed in a 5-d
bulk, but exclude changes of signature a priori, they would take
the line element (\ref{oldRW}) and the associated cosmic time $T$
to describe all the Universe and its history.
Moreover, the metric volume form would be chosen
to normalise the energy-momentum tensor $\tau_{\mu\nu}$
as well as $\hat\varrho$
and $\hat p$. Denoting them simply by $\varrho$ and $p$,  
their explicit expressions read (Mars \etal \/ \cite{signlong})

\[
\varrho'+3\frac{a'}{a}({\varrho}+{p})=0,~~~
\label{eq:cons}
  \\
\frac{\kappa^2_5}{3}\, \varrho=
  \sigma\epsilon_1
  \left(
    \sqrt{\frac{{a'}^2+k}{a^2}
      +\Blambda^2 }
    - \sqrt{\frac{{a'}^2+k}{a^2}
      +\lambda^2}\,
  \right),
  \label{eq:Friedmann_L}
\]
where $'\equiv d/dT$, $\kappa_{5}^2$ is the 5-dimensional
gravitational coupling constant
and $\epsilon_{1}$ is a sign selecting which region bounded
by $\S$ in AdS$_{5}$ is to be matched to which region bounded by
$\tilde\S$ in $\widetilde{\mbox{AdS}}_5$. 
The metric volume form on the Lorentzian phase 
obviously becomes singular on the signature changing set 
$S\cap\overline{\S_L}$. This may seem to imply that ${\varrho}$ and ${p}$
also have to diverge necessarily there.
However, this is not the case and, in fact, 
${\varrho}$ {\em must vanish} at the signature change,
and  $p$ can also be regular there
in many situations (Mars \etal \/ \cite{signlong}).
In fact, there exists a universal bound for the 
energy-density $\rho$ in the Lorentzian phase given by
$\frac{\kappa_5^2}{3} |\varrho| \leq  \sqrt{|\Blambda^2-\lambda^2|}$.

Consequently, appropriate choices of 
hypersurfaces in AdS$_{5}$
and $\widetilde{\mbox{AdS}}_5$ allow one to {\em construct 
signature-changing branes with both $\varrho$ and $p$ finite and
well-behaved everywhere on $\overline{\S_L}$}. As a matter of fact,
signature-changing branes satisfying some desirable energy conditions
can be built, as we show next.
%

\section{Example}
This model is based on the equation of state (on $\S_L$)
$
p=C^2 \varrho^{\frac{m-2}{m}}
$
with $m$ odd. The conservation
equation (\ref{eq:cons}) implies 
$\varrho=C^{m}\left[\left(\frac{a_S}{a}\right)^{6/m}-1\right]^{m/2}$,
where $a_S>0$ is an integration constant. 
In order to construct a regular brane
$\S$ we need
$\ff(\xi) =(\xi-\xi_e)^{m}(\xi-\xi_b)^{m}$, so that 
the Lorentzian phase $\S_L$ corresponds to 
$\xi_b<\xi<\xi_e$. The function $a(\xi)$ satisfies an ODE which admits the
boundary
conditions $a(\xi_b)=a(\xi_e)=a_S$. The corresponding solution can be proven
(Mars \etal \/ \cite{sudden})
to be \emph{regular}
all over $\S$ and satisfy  $a_{min}\leq a\leq a_S$ on $\S_L$, which implies
in particular that the weak and strong energy conditions are satisfied.
Moreover, $a'(T)\to \pm\infty$ and $a''(T)\to \infty $ as $a\to a_S$
and the model describes a sudden type singularity with a diverging
Hubble factor for a finite non-zero scale factor $a=a_S$. The brane
can be extended past $\xi_{e}$ and $\xi_b$ smoothly so that this 
apparent singularity
is just an artifact of the bad behaviour of $T$,  
as discussed above. 


\begin{figure}[h]
  \centering
  \parbox[c]{6cm}{\includegraphics[width=4.6cm]{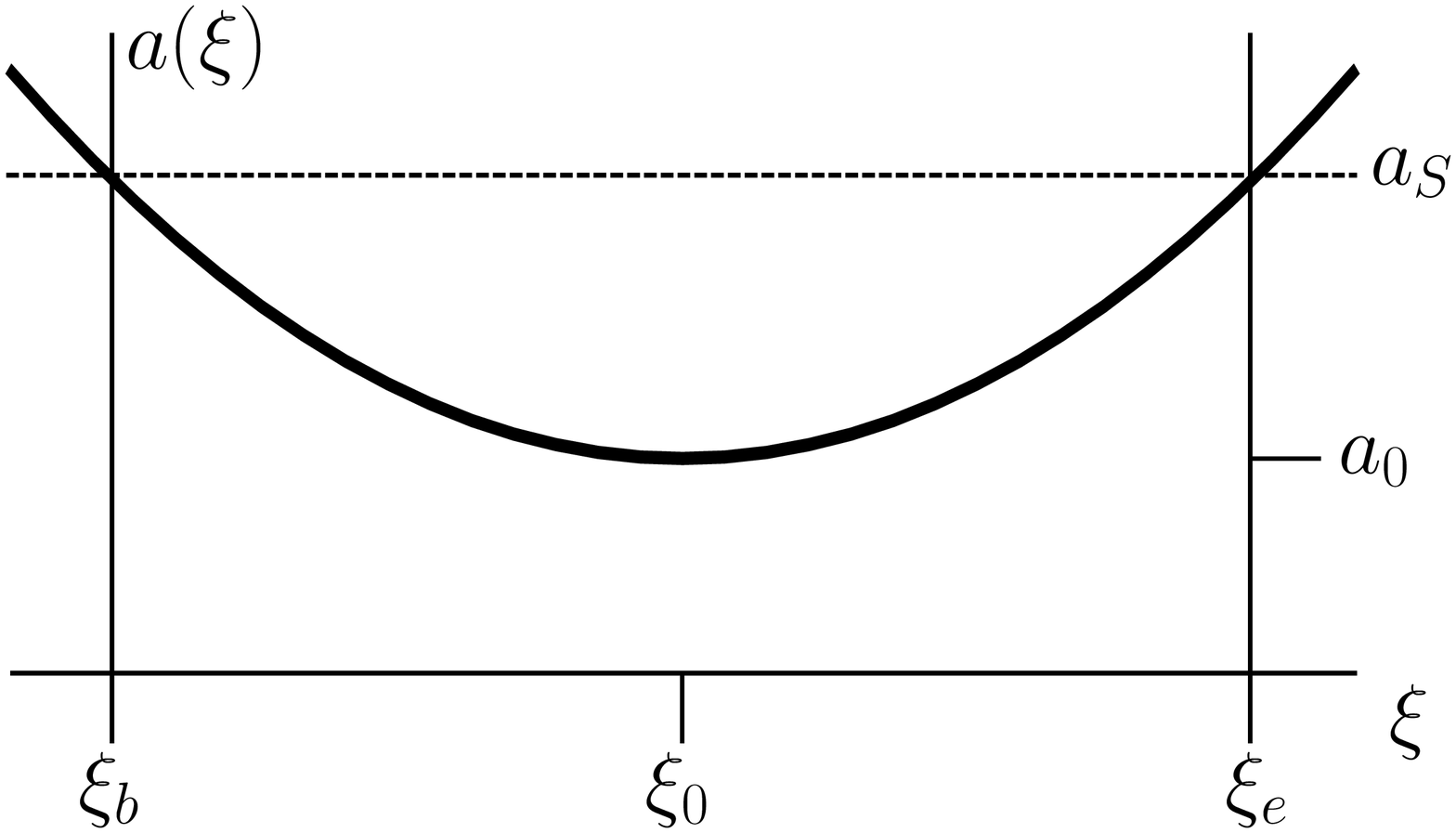}
      \includegraphics[width=4cm]{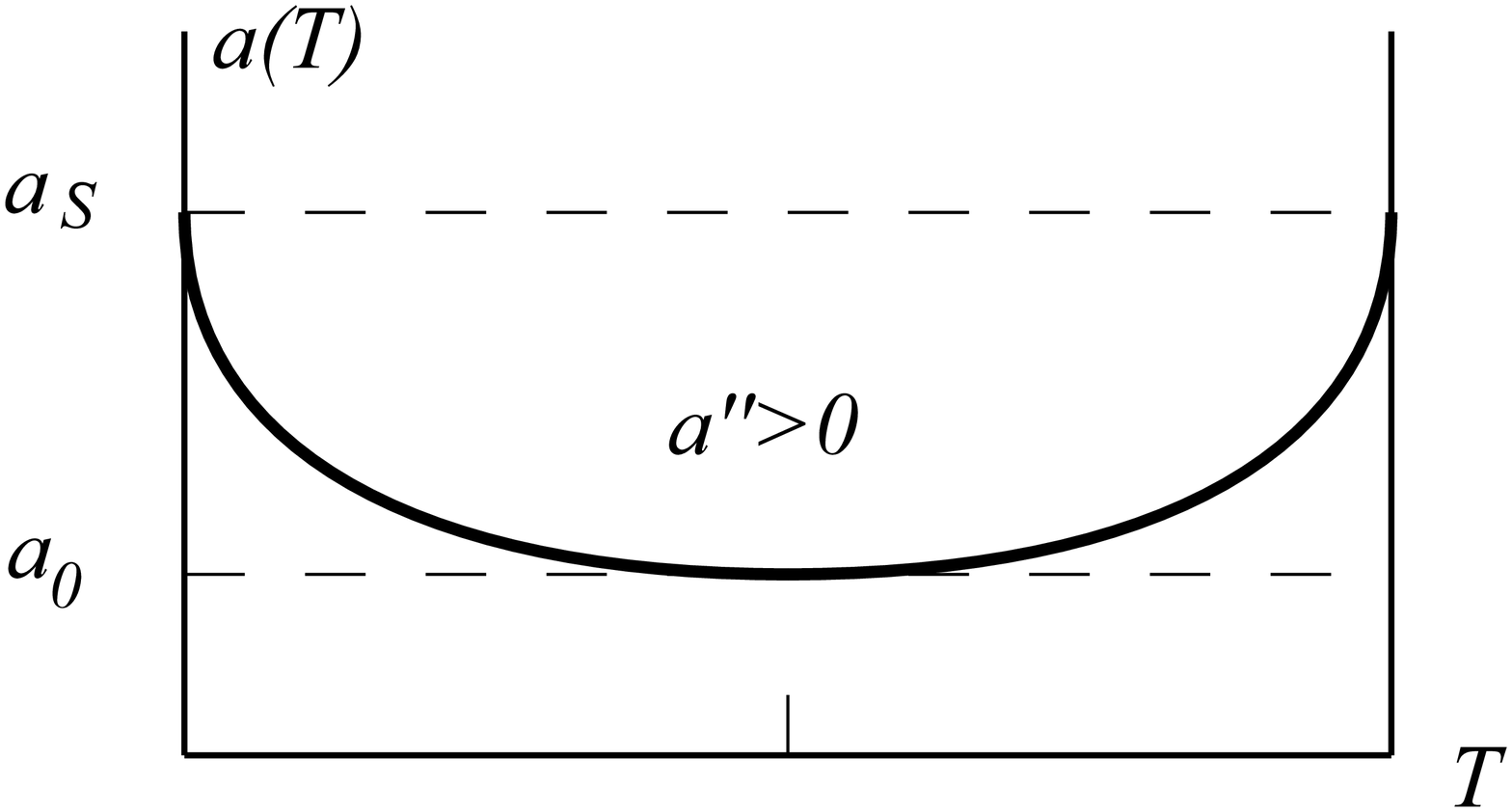}}%
  \parbox[c]{4cm}{\includegraphics[width=3.9cm]{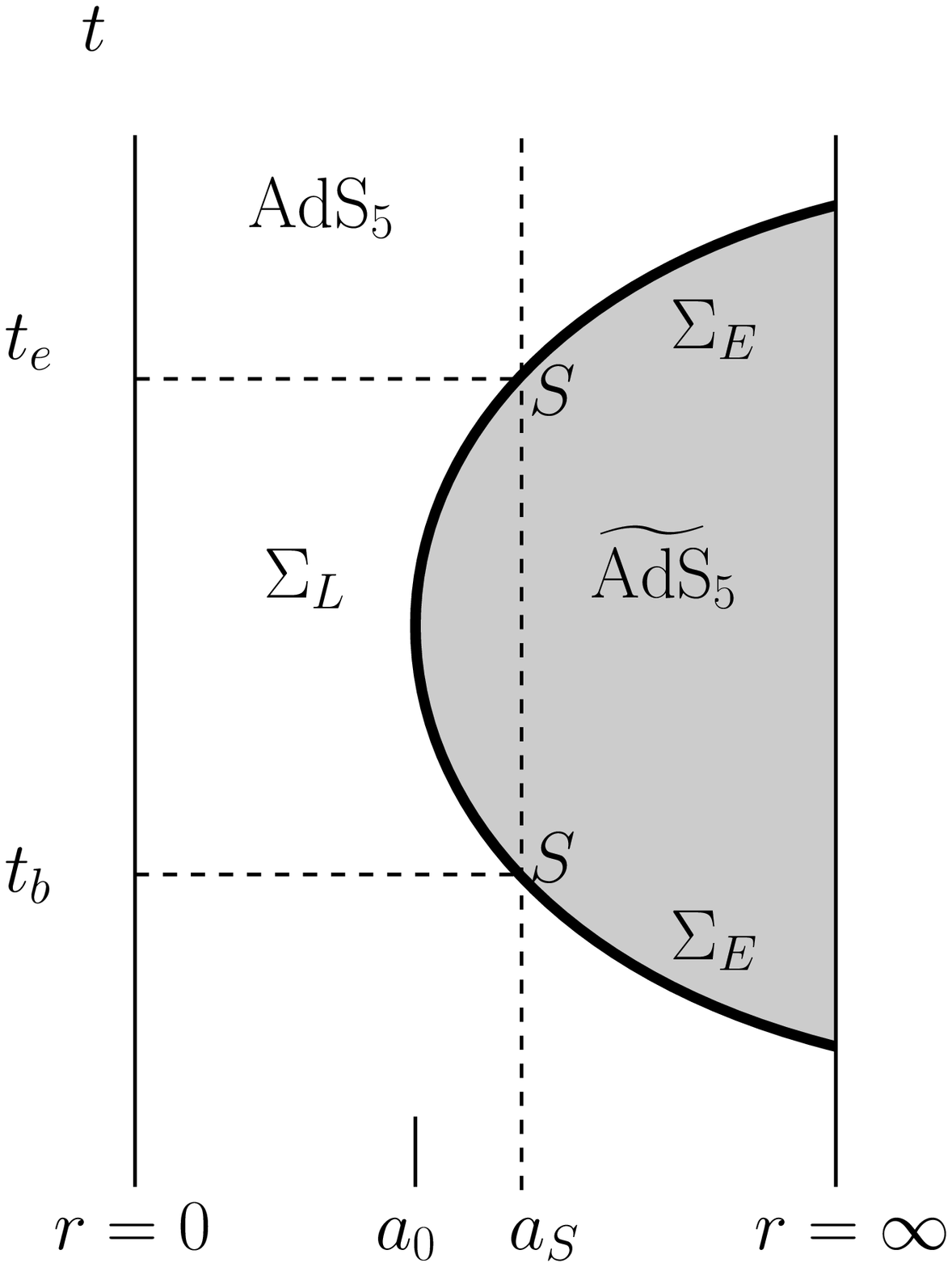}}
  \caption{$\S_L$ starts
    at a change of signature at $\xi_b$ where/when the cosmic time
    $T$ is born. There, the scale factor $a(T)$
    has a maximum and  $\varrho$ and $p$ vanish.
    From there, $a(T)$ contracts to a minimum, $a_0$,
    where $\varrho$ attains its maximum (regular \emph{little bang}).
    From this bounce, the model expands with an accelerated rate and $T$
    ends in a \emph{seemingly} sudden singularity, at the signature
    change at $\xi_e$, where again $\varrho=p=0$.
    The picture on the right describes the whole brane embedded
    in the AdS$_5$ ($k=1$) bulks, with $\sigma=-\epsilon_1=1$, and
    $\lambda>\Blambda$, so that $\varrho>0$ on $\S_L$.
  }
\label{fig:main}
\end{figure}


\end{document}